\documentclass[final,5p,times]{elsarticle}

\usepackage[T1]{fontenc}
\usepackage[utf8]{inputenc}
\usepackage[english]{babel}
\usepackage{amsmath}
\usepackage{amssymb}
\usepackage{lipsum}
\usepackage{hyperref}
\usepackage{adjustbox}
\usepackage{color}
\usepackage[percent]{overpic}
\usepackage{float}
\usepackage[caption = false]{subfig}
\usepackage{lineno}
\usepackage{orcidlink} 
\usepackage{etoolbox}

\allowdisplaybreaks

\makeatletter
\def\ps@pprintTitle{%
  \let\@oddhead\@empty
  \let\@evenhead\@empty
  \def\@oddfoot{\reset@font\hfil\thepage\hfil}
  \let\@evenfoot\@oddfoot
}
\makeatother

\begin{document}

\begin{frontmatter}


\title{Measurement of Spin-Density Matrix Elements in $\Delta^{++}(1232)$ photoproduction}
\affiliation{organization={Polytechnic Sciences and Mathematics, School of Applied Sciences and Arts, Arizona State University}, city={Tempe}, state={Arizona}, postcode={85287}, country={USA}}
\affiliation{organization={Department of Physics, National and Kapodistrian University of Athens}, postcode={15771}, city={Athens}, country={Greece}}
\affiliation{organization={Ruhr-Universität-Bochum, Institut für Experimentalphysik}, postcode={D-44801}, city={Bochum}, country={Germany}}
\affiliation{organization={Helmholtz-Institut für Strahlen- und Kernphysik, Universität Bonn}, postcode={D-53115}, city={Bonn}, country={Germany}}
\affiliation{organization={Department of Physics, Carnegie Mellon University}, city={Pittsburgh}, state={Pennsylvania}, postcode={15213}, country={USA}}
\affiliation{organization={Department of Physics The Catholic University of America}, city={Washington, D.C}, postcode={20064}, country={USA}}
\affiliation{organization={Department of Physics, University of Connecticut}, city={Storrs}, state={Connecticut}, postcode={06269}, country={USA}}
\affiliation{organization={Department of Physics, Duke University}, city={Durham}, state={North Carolina}, postcode={27708}, country={USA}}
\affiliation{organization={Department of Physics, Florida International University}, city={Miami}, state={Florida}, postcode={33199}, country={USA}}
\affiliation{organization={Department of Physics, Florida State University}, city={Tallahassee}, state={Florida}, postcode={32306}, country={USA}}
\affiliation{organization={Department of Physics, The George Washington University}, city={Washington, D.C.}, postcode={20052}, country={USA}}
\affiliation{organization={School of Physics and Astronomy, University of Glasgow}, city={Glasgow}, postcode={G12 8QQ}, country={United Kingdom}}
\affiliation{organization={GSI Helmholtzzentrum für Schwerionenforschung GmbH}, postcode={D-64291}, city={Darmstadt}, country={Germany}}
\affiliation{organization={Institute of High Energy Physics}, city={Beijing}, postcode={100049}, country={People's Republic of China}}
\affiliation{organization={Department of Physics, Indiana University}, city={Bloomington}, state={Indiana}, postcode={47405}, country={USA}}
\affiliation{organization={National Research Centre Kurchatov Institute}, city={Moscow}, postcode={123182}, country={Russia}}
\affiliation{organization={Department of Physics, Lamar University}, city={Beaumont}, state={Texas}, postcode={77710}, country={USA}}
\affiliation{organization={Department of Physics, University of Massachusetts}, city={Amherst}, state={Massachusetts}, postcode={01003}, country={USA}}
\affiliation{organization={National Research Nuclear University Moscow Engineering Physics Institute}, city={Moscow}, postcode={115409}, country={Russia}}
\affiliation{organization={Department of Physics, Mount Allison University}, city={Sackville}, state={New Brunswick}, postcode={E4L 1E6}, country={Canada}}
\affiliation{organization={Department of Physics, Norfolk State University}, city={Norfolk}, state={Virginia}, postcode={23504}, country={USA}}
\affiliation{organization={Department of Physics, North Carolina A\&T State University}, city={Greensboro}, state={North Carolina}, postcode={27411}, country={USA}}
\affiliation{organization={Department of Physics and Physical Oceanography}, city={University of North Carolina at Wilmington, Wilmington}, North Carolina postcode={28403}, country={USA}}
\affiliation{organization={Department of Physics, Old Dominion University}, city={Norfolk}, state={Virginia}, postcode={23529}, country={USA}}
\affiliation{organization={Department of Physics, University of Regina}, city={Regina}, state={Saskatchewan}, postcode={S4S 0A2}, country={Canada}}
\affiliation{organization={Department of Mathematics, Physics, and Computer Science, Springfield College}, city={Springfield}, state={Massachusetts}, postcode={01109}, country={USA}}
\affiliation{organization={Thomas Jefferson National Accelerator Facility}, city={Newport News}, state={Virginia}, postcode={23606}, country={USA}}
\affiliation{organization={Laboratory of Particle Physics, Tomsk Polytechnic University}, postcode={634050}, city={Tomsk}, country={Russia}}
\affiliation{organization={Department of Physics, Tomsk State University}, postcode={634050}, city={Tomsk}, country={Russia}}
\affiliation{organization={Department of Physics and Astronomy, Union College}, city={Schenectady}, state={New York}, postcode={12308}, country={USA}}
\affiliation{organization={Department of Physics, Virginia Tech}, city={Blacksburg}, state={VA}, postcode={24061}, country={USA}}
\affiliation{organization={Department of Physics, Washington \& Jefferson College}, city={Washington}, state={Pennsylvania}, postcode={15301}, country={USA}}
\affiliation{organization={Department of Physics, William \& Mary}, city={Williamsburg}, state={Virginia}, postcode={23185}, country={USA}}
\affiliation{organization={School of Physics and Technology, Wuhan University}, city={Wuhan}, state={Hubei}, postcode={430072}, country={People's Republic of China}}
\affiliation{organization={A. I. Alikhanyan National Science Laboratory (Yerevan Physics Institute)}, postcode={0036}, city={Yerevan}, country={Armenia}}
\author{F.~Afzal\orcidlink{0000-0001-8063-6719}$^\textrm{\scriptsize d}$\corref{cor1}}
\cortext[cor1]{afzal@jlab.org}
\author{C.~S.~Akondi\orcidlink{0000-0001-6303-5217}$^\textrm{\scriptsize j}$}
\author{M.~Albrecht\orcidlink{0000-0001-6180-4297}$^\textrm{\scriptsize aa}$}
\author{M.~Amaryan\orcidlink{0000-0002-5648-0256}$^\textrm{\scriptsize x}$}
\author{S.~Arrigo$^\textrm{\scriptsize ag}$}
\author{V.~Arroyave$^\textrm{\scriptsize i}$}
\author{A.~Asaturyan\orcidlink{0000-0002-8105-913X}$^\textrm{\scriptsize aa}$}
\author{A.~Austregesilo\orcidlink{0000-0002-9291-4429}$^\textrm{\scriptsize aa}$}
\author{Z.~Baldwin\orcidlink{0000-0002-8534-0922}$^\textrm{\scriptsize e}$}
\author{F.~Barbosa$^\textrm{\scriptsize aa}$}
\author{J.~Barlow\orcidlink{0000-0003-0865-0529}$^\textrm{\scriptsize j,z}$}
% ...

\begin{abstract}
We measure the spin-density matrix elements (SDMEs) of the $\Delta^{++}(1232)$ in the photoproduction reaction $\gamma p \to \pi^-\Delta^{++}(1232)$ with the GlueX experiment in Hall D at Jefferson Lab. The measurement uses a linearly--polarized photon beam with energies from $8.2$ to $8.8$~GeV and the statistical precision of the SDMEs exceeds the previous measurement by three orders of magnitude for the momentum transfer squared region below $1.4$ GeV$^2$. The data are sensitive to the previously undetermined relative sign between couplings in existing Regge-exchange models. Linear combinations of the extracted SDMEs allow for a decomposition into natural and unnatural--exchange amplitudes. We find that the unnatural exchange plays an important role in the low momentum transfer region.   
\end{abstract}





\newpageafter{author}

\end{frontmatter}

\section{Introduction}
\label{introduction}
In recent years, there have been many discoveries of exotic hadrons containing both light and heavy quarks that cannot be described as conventional hadrons (mesons or baryons) in the quark model, see \textit{e.g.} Refs.~\cite{PhysRevLett.91.262001,PhysRevLett.104.241803,PhysRevLett.115.072001,MEYER201521}.
The GlueX experiment at Jefferson Lab studies the photoproduction of light-quark mesons and baryons, with an emphasis on the search for one type of these exotic hadrons, namely hybrid mesons. These hybrid mesons consist of an excited gluonic field coupled to a quark-antiquark pair, which contributes to the quantum numbers of the mesons, allowing for exotic quantum numbers ($J^{PC}=0^{--}, 0^{+-}, 1^{-+},
2^{+-}, \dots)$. At high beam energies ($E_\gamma\approx8.5$~GeV), photoproduction with a linearly-polarized photon beam can be described by $t$-channel Regge exchanges with contributions from natural-parity ($\eta=+1$, \textit{e.g.} $\rho$) and unnatural-parity ($\eta=-1$, \textit{e.g.} $\pi$) exchange. The naturality $\eta=P(-1)^J$ is defined by the parity $P$ and total angular momentum $J$ of the exchanged Reggeon. When studying the meson spectrum through photoproduction, understanding these exchange mechanisms is important for the development of the amplitude analyses required to study known states and search for new states.

The reaction $\gamma p \rightarrow \pi^-\Delta^{++}$ offers the opportunity to investigate the charge-exchange production mechanism, in particular, the unnatural-parity pion exchange.  We report on the measurement of spin-density matrix elements (SDMEs) of the $\Delta^{++}(1232)$ in the reaction $\gamma p \to \pi^-\Delta^{++}(1232)$ with a linearly-polarized photon beam with an average beam energy of $E_\gamma=8.5$~GeV.  These SDMEs describe the spin polarization of the spin-$\frac{3}{2}$ $\Delta^{++}$ in terms of the underlying helicity amplitudes and can be used to separate contributions from the different parity exchanges.

Theoretical models have been developed by the Joint Physics Analysis Center (JPAC)~\cite{nysFeaturesPiDelta2018} and Yu and Kong~\cite{YU2017262} to describe $\pi\Delta$ photoproduction using Regge theory amplitudes. These models are constrained by previous measurements of the dependence on the four-momentum transfer squared $t=(p_\gamma - p_{\pi^-})^2$ of both the differential cross section~\cite{PhysRevLett.22.148,PhysRevD.20.1553} and the linearly-polarized beam asymmetry~\cite{PhysRevD.20.1553,PhysRevC.103.L022201}, where $p_\gamma$ and $p_{\pi^-}$ are the four-momenta of the beam photon and the $\pi^-$. There is only one previous measurement of the $\Delta^{++}(1232)$ SDMEs, which had limited statistical precision and coverage in $t$~\cite{PhysRevD.7.3150}. The measurement presented here provides the first determination of the $t$-dependence of the SDMEs, which confirms the importance of unnatural-parity exchange in the low $-t$ region. This exchange is not well described by existing models, demonstrating the lack of constraints provided by previous measurements.

\section{The GlueX experiment}
\label{setup}
The data presented here were collected by the GlueX experiment, located in Hall D at Jefferson Lab. The Continuous Electron Beam Accelerator Facility (CEBAF) provides an electron beam of 11.6 GeV which is incident on a thin diamond crystal of $50~\mu$m thickness, producing linearly-polarized photons via coherent bremsstrahlung. The coherent peak is placed at a beam photon energy of $E_\gamma=8.8$~GeV, giving a polarization degree of $28$ to $36$\% in the $E_\gamma$ range between $8.2$ and $8.8$~GeV. The degree of linear polarization is measured with a triplet polarimeter \cite{DUGGER2017115} and has an estimated systematic uncertainty of 1.5\%. To control for systematic effects, the diamond orientation is changed regularly between two pairs of orthogonal linear polarization angles: $0^\circ, 90^\circ$ and $45^\circ, -45^\circ$. A tagging system consisting of a dipole magnet and scintillation detectors placed in the tagger focal plane measures the scattered electrons' momenta, determining the energy and production time of each bremsstrahlung photon.

The photon beam impinges on a 30 cm-long liquid hydrogen target, surrounded by a detector system consisting of the Start Counter (SC) \cite{Pooser:2019rhu}, the Central \cite{Jarvis:2019mgr} and Forward \cite{Pentchev:2017omk} Drift Chambers (CDC and FDC), as well as the Barrel \cite{Beattie:2018xsk} and Forward Calorimeters (BCAL and FCAL) and the forward time-of-flight (TOF) detector. A superconducting solenoid with an average field strength of 2~T surrounds the central detectors. Charged particles are tracked in the CDC and FDC, providing a momentum resolution of $1$ to $5$\%. In addition, the time-of-flight measurement in the BCAL and TOF and the measurement of the energy loss $dE/dx$ in the CDC are used for particle identification. More details about the setup are given in Ref.~\cite{ADHIKARI2021164807}.

The results presented here are obtained from the first phase of the GlueX experiment, collected in 2017 and 2018, with a total integrated luminosity of about 125 pb$^{-1}$ in the coherent peak region.

\section{Spin-density matrix elements}
\subsection{Event Selection}
\label{sec:eventsel}
The reaction $\gamma p \to \pi^-\Delta^{++}(1232)$ is studied, where the $\Delta^{++}(1232)$ decays to $\pi^+p$. Exclusive events with the $\pi^-\pi^+p$ final-state topology are selected by imposing several requirements on the data. Events with at least two positively charged particles and one negatively charged particle are used for the analysis and each possible combination of these particles is analyzed. Events with up to three additional charged tracks are allowed to avoid rejecting signal events with spurious uncorrelated (or random) tracks. The proton and charged-pion candidates are identified by time-of-flight requirements using the charged-particle timing information from the SC, BCAL, FCAL, and TOF detectors and the energy loss $dE/dx$ in the CDC.

Incident beam photons are required to have an energy in the coherent peak, \textit{i.e.} $8.2<E_\gamma <8.8$~GeV, as measured by the tagger. Using the total initial- and final-state four-momenta $p_i$ and $p_f$, events with the measured missing mass squared of $m^2_{\text{miss}}=|p_i - p_f|^2 < 0.1$~GeV$^2$ are selected, thus removing events where unwanted massive particles are produced but not detected. In addition, a kinematic fit to the reaction hypothesis $\gamma p \to \pi^-\pi^+p$ is applied for each event, imposing energy and momentum conservation, as well as a common vertex for all particles. Events  fulfilling a kinematic fit quality criterion of $\chi^2/\text{ndf}~<~8.7$ are retained.

The initial-state beam photon and the final-state particles should be coincident in time. The electron beam is provided by the CEBAF accelerator in beam bunches 4 ns apart, with the precise bunch time given by the accelerator radiofrequency (RF) clock. Due to the high intensity of the electron beam, the tagger can detect several photons from the same beam bunch, while only one of these photons interacts with the liquid hydrogen target. The time difference between the final-state particles and the tagged beam bunch is required to be less than 2~ns. To account for accidental tagger hits, four beam bunches before and after the prompt peak signal are used with a weight of $-\frac{1}{8}$ to subtract the accidental in-time background within the time window of the signal. 

\begin{figure}
    \centering 
    \begin{overpic}[width=\columnwidth]{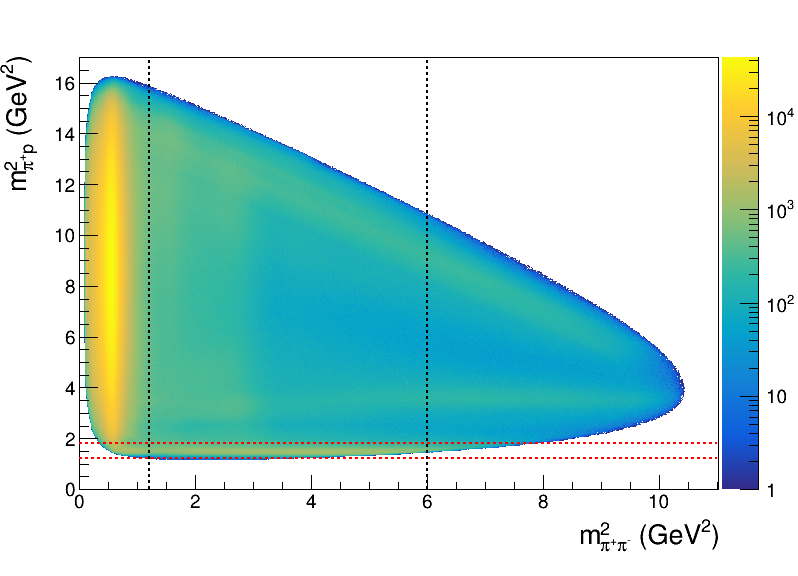}
    \put(60,15){\scriptsize{$\leftarrow\Delta^{++}(1232)$}}
\put(70,32){\makebox(0,0){\rotatebox{-25}{\scriptsize{$\leftarrow \Delta^0/N^*$}}}}
\put(14,56){\makebox(0,0){\rotatebox{90}{\scriptsize{$\leftarrow \rho(770)$}}}}
\end{overpic}
    \caption{Measured Dalitz plot of the reaction $\gamma p \to \pi^-\pi^+p$. Vertical bands show contributions from the $\pi^-\pi^+$ meson system \textit{e.g.} $\rho(770)$, while horizontal bands show the $\Delta^{++}(1232)$ and excited baryon states that decay to $\pi^+p$. On the diagonal, baryon contributions from $N^*/\Delta^{0*}$ are visible in the $\pi^-p$ system. The dashed black and red lines show the chosen selections on the $\pi^-\pi^+$ and $\pi^+p$ masses, respectively.}
    \label{fig:Dalitz}
\end{figure}

The Dalitz plot for this channel is shown in Fig.~\ref{fig:Dalitz}. The $\Delta^{++}(1232)$ is selected by requiring the $\pi^+p$ invariant mass of $1.10< m_{\pi^+p}<1.35$~GeV. The upper limit is chosen to avoid significant contribution from higher-mass excited $\Delta^*$ states. Permutations of the final-state particles $\pi^-\pi^+p$, can lead to background contributions from the $\pi^-p$ baryon system or $\pi^-\pi^+$ meson system. The $\pi^-p$ baryon system, consisting of $N^*/\Delta^{0*}$ baryon resonances, is seen on the diagonal of the Dalitz plot in Fig.~\ref{fig:Dalitz}. Restricting the analysis to the range of four-momentum transfer squared below $1.4$~GeV$^2$ removes most of this background. In addition, this background is well separated from the selected $\Delta^{++}(1232)$ mass region in $m_{\pi^+p}$ and can therefore be neglected. However, the contribution from the $\pi^-\pi^+$ meson system poses a significant background, as shown in Fig.~\ref{fig:Dalitz}, where the $\rho(770)$ dominates. The $\rho(770)$ contribution can be significantly reduced by restricting the $\pi^-\pi^+$ mass range to $1.10< m_{\pi^-\pi^+}<2.45$~GeV. The upper limit reduces the contribution from excited $\Delta^*$. Despite the applied selection of the $\pi^-\pi^+$ mass, a significant irreducible amount of background remains that needs to be taken into consideration. 

\begin{figure}[h!]
\centering 
    \subfloat[]{
    	\begin{overpic}[width=0.49\columnwidth]{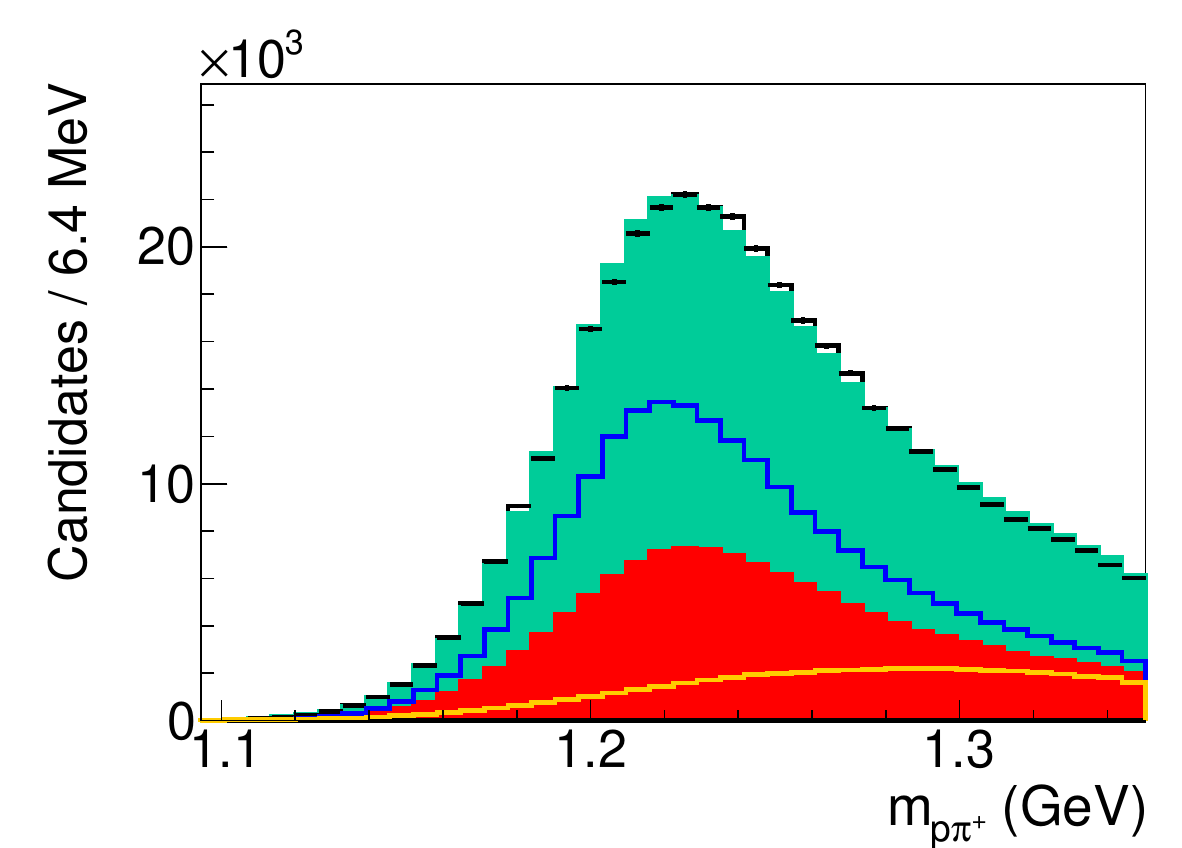}	
            \put(27,70){\scriptsize{$0.100$~GeV$^2<|t|\leq 0.125$~GeV$^2$ }}
        \end{overpic}
        \label{subfig:DeltaMass_tbin2}
    }
    \subfloat[]{
        \begin{overpic}[width=0.49\columnwidth]{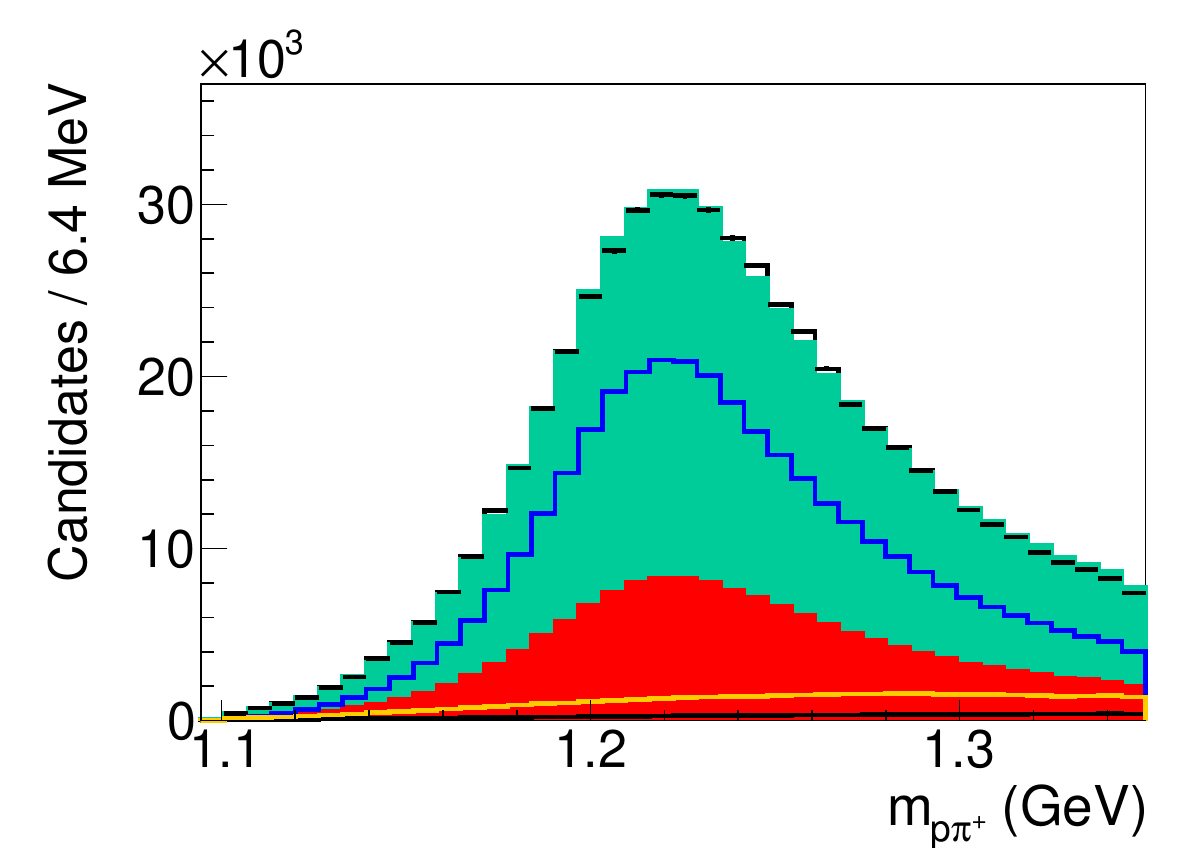}	
            \put(27,70){\scriptsize{$0.640$~GeV$^2<|t|\leq 0.850$~GeV$^2$ }}
            \label{subfig:DeltaMass_tbin11}
        \end{overpic}
    }\\
    \subfloat[]{
        \includegraphics[width=0.49\columnwidth]{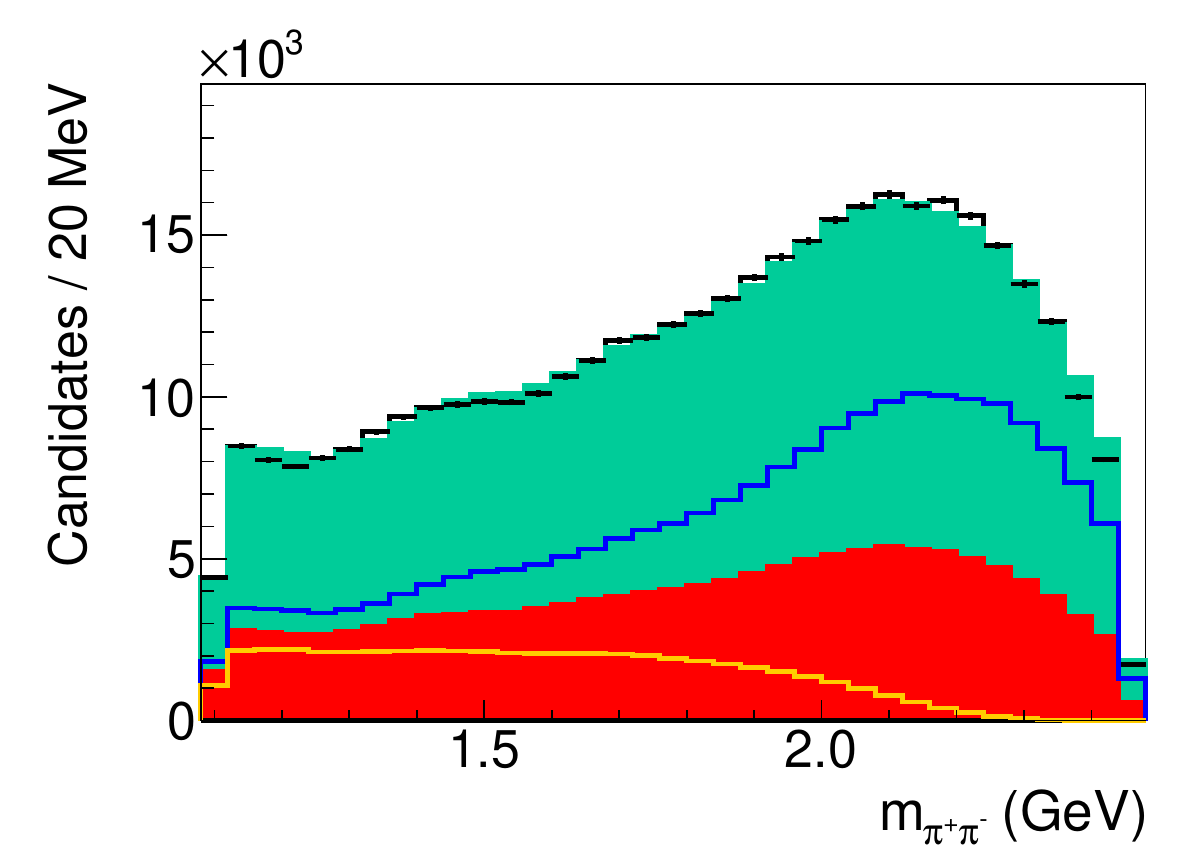}	
        \label{subfig:2PiMass_tbin2}
    }
    \subfloat[]{
        \includegraphics[width=0.49\columnwidth]{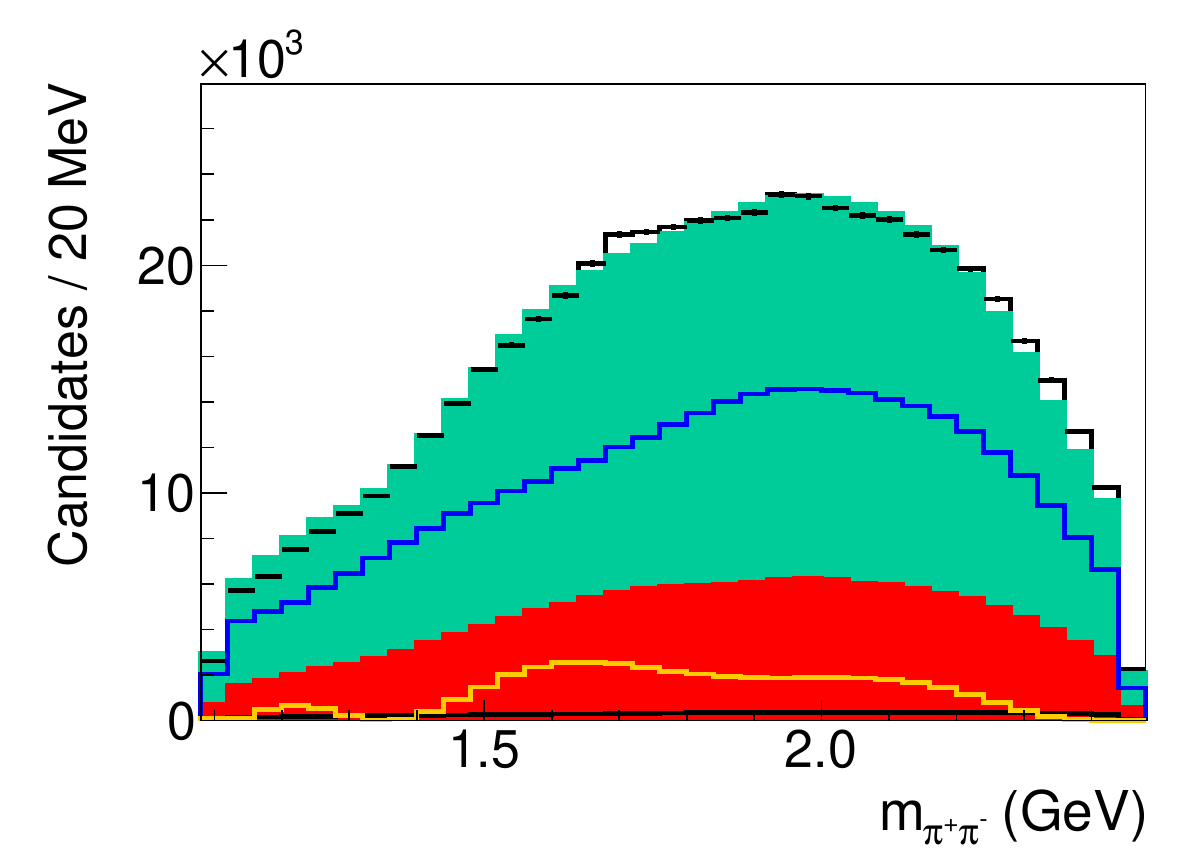}
        \label{subfig:2PiMass_tbin11}
    }\\
    \subfloat[]{
        \includegraphics[width=0.49\columnwidth]{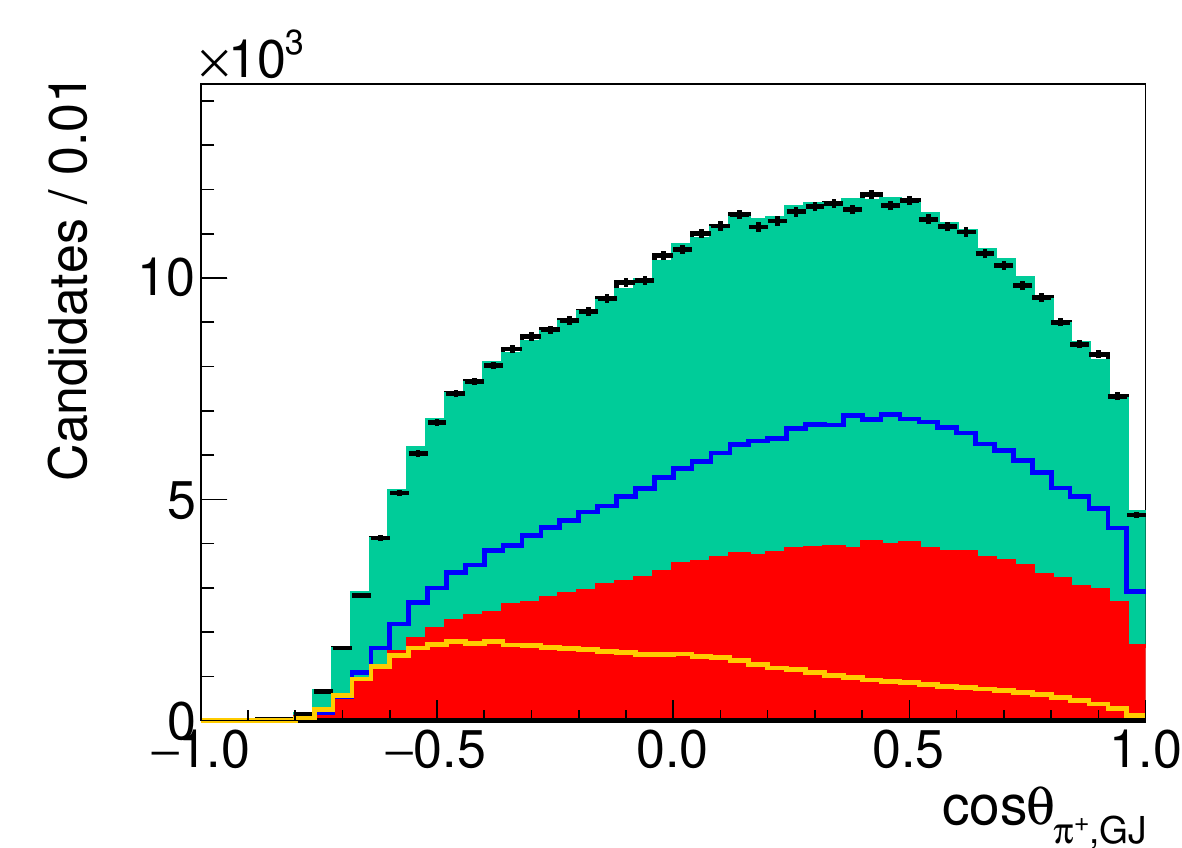}
        \label{subfig:cosT_tbin2}
    }
    \subfloat[]{
        \includegraphics[width=0.49\columnwidth]{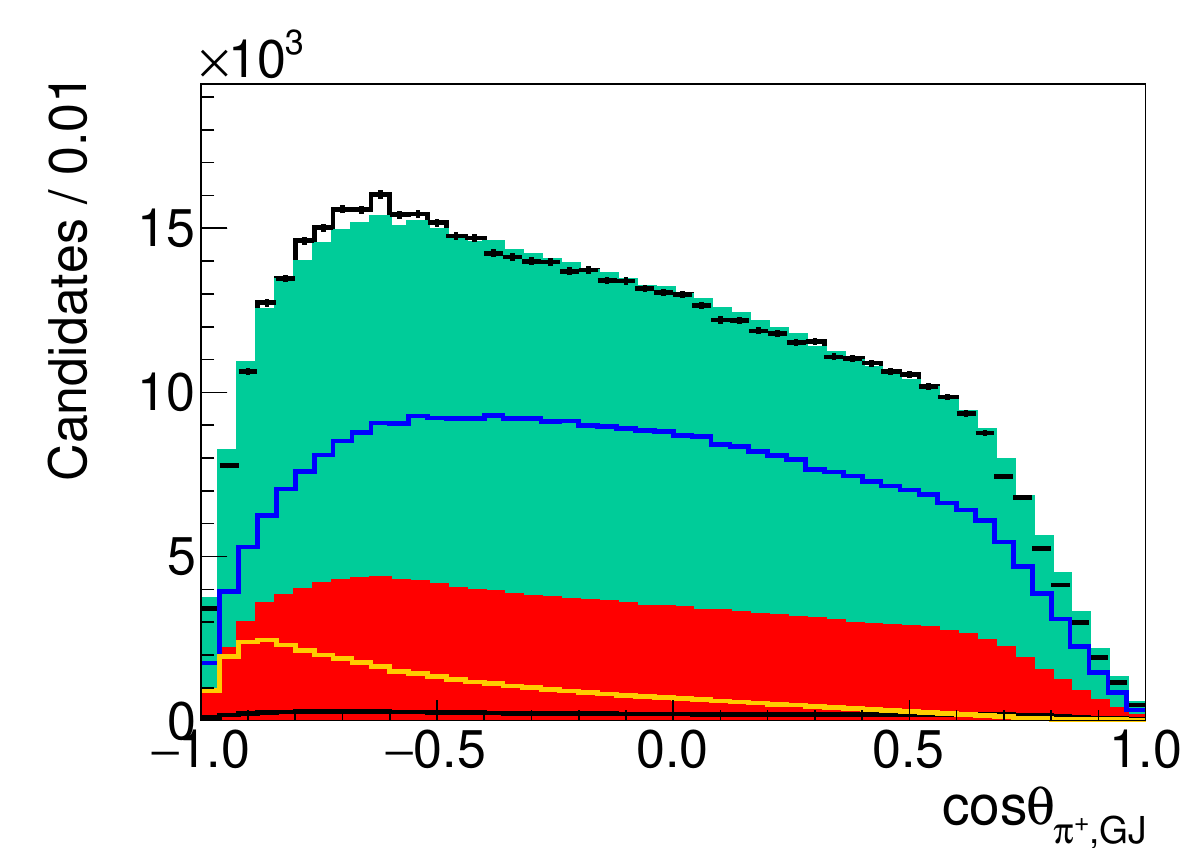}
        \label{subfig:cosT_tbin11}
    }\\
    \subfloat[]{
        \includegraphics[width=0.49\columnwidth]{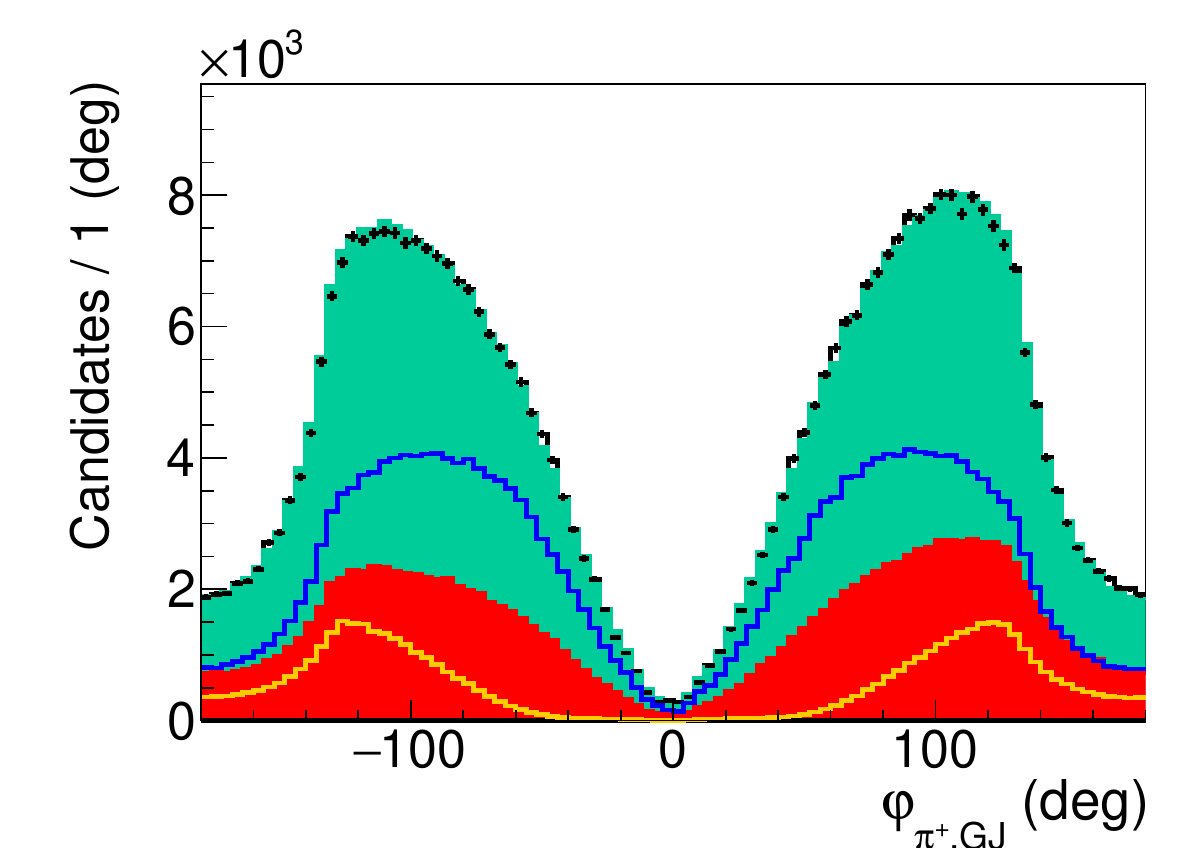}	
        \label{subfig:phi_tbin2}
    }
    \subfloat[]{
        \includegraphics[width=0.49\columnwidth]{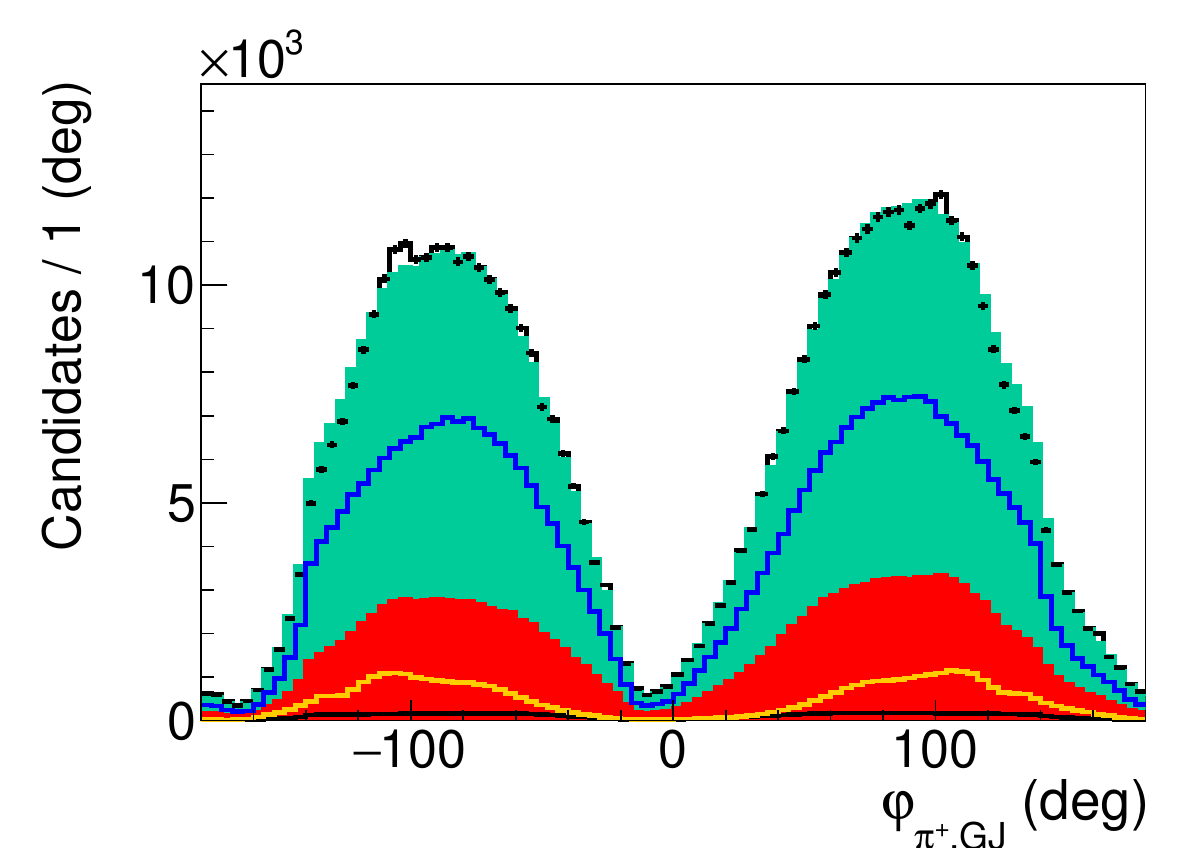} 
        \label{subfig:phi_tbin11}
    }	\\
    \subfloat[]{
        \includegraphics[width=0.49\columnwidth]{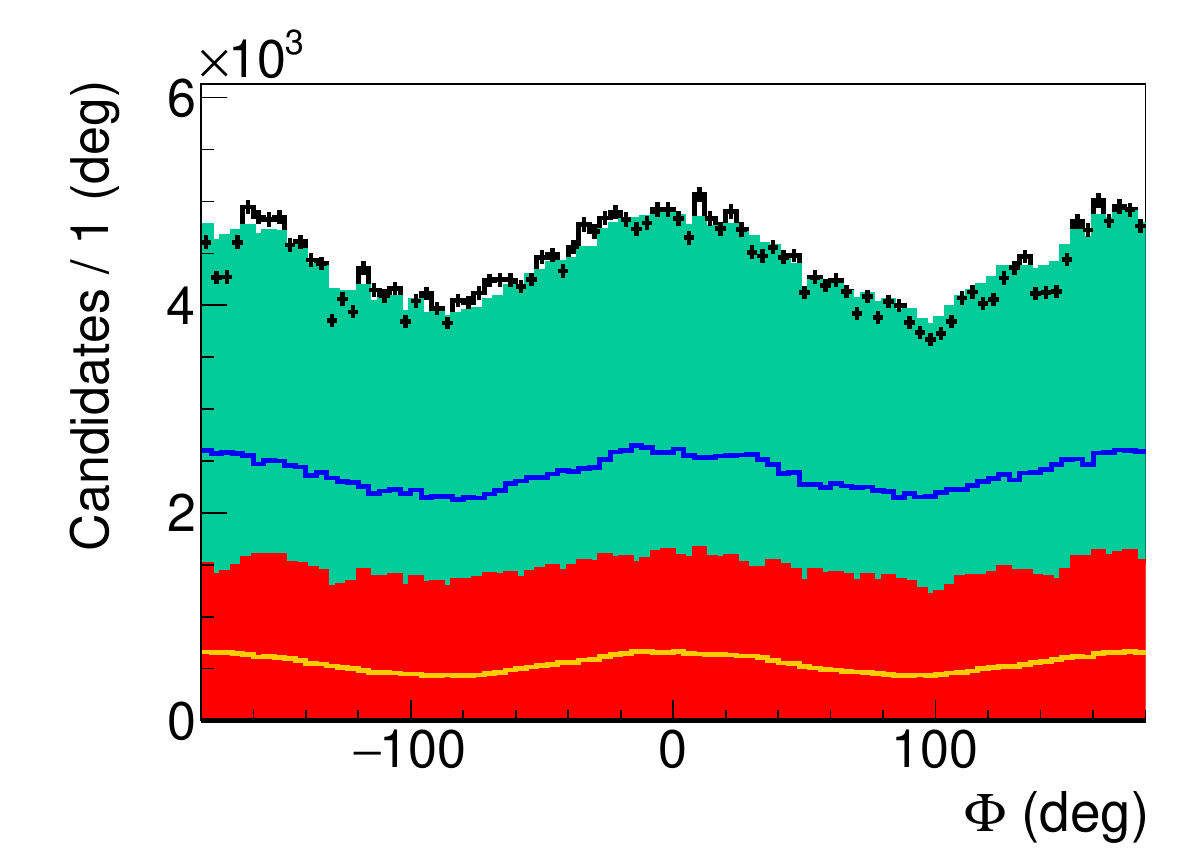}	
        \label{subfig:Phi_tbin2}
    }
    \subfloat[]{
        \includegraphics[width=0.49\columnwidth]{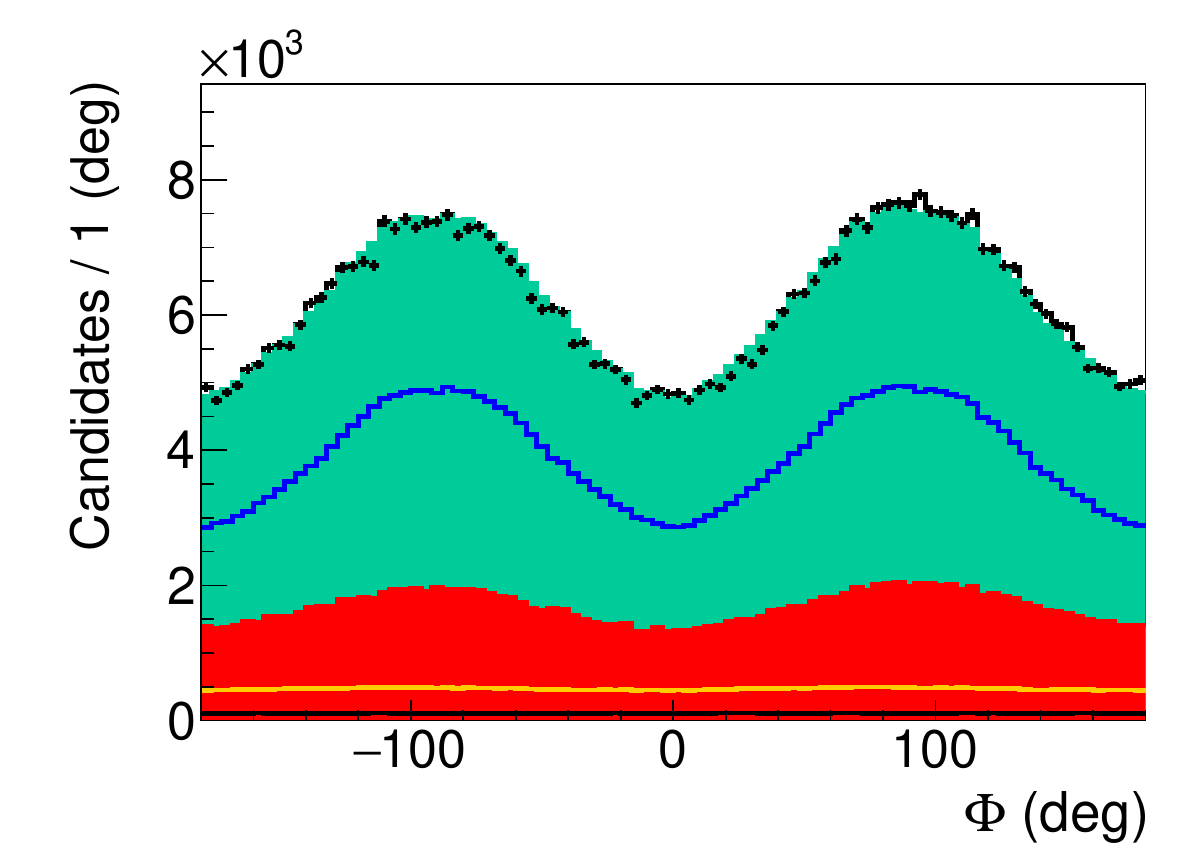}	
        \label{subfig:Phi_tbin11}
    }
    \caption{
    Measured distributions (black points) compared to simulated phase-space distributions weighted by the detector acceptance and the fit results (green shaded area). The accidental time background distribution is depicted as the red shaded area. The blue, yellow, and black lines show the $\pi^-\Delta^{++}$, $(\pi^-\pi^+) p$ and phase space contributions, respectively, in the fit model given by Eq.~\ref{eq:fitintensity}. The left and right columns show results for two representative $t$ bins. The phase space contribution is very small.
    }
	\label{fig_eventsel}%
\end{figure}

Figure~\ref{fig_eventsel} shows the mass spectra for two example $-t$ bins: 
The black points show the $\pi^+p$ mass (see Figs.~\ref{subfig:DeltaMass_tbin2} and ~\ref{subfig:DeltaMass_tbin11}) and the $\pi^-\pi^+$ mass (see Figs.~\ref{subfig:2PiMass_tbin2} and ~\ref{subfig:2PiMass_tbin11}) distributions. The red shaded areas show the accidental time background contribution which makes up roughly 30\% of the selected data sample. The $m_{\pi^-\pi^+}$ dependence of the background is parameterized by a Bernstein polynomial of 4th degree (see Eq.~\ref{eq:Bernstein}), as shown by the yellow curves in the $m_{\pi^-\pi^+}$ distributions of Figs.~\ref{subfig:2PiMass_tbin2} and ~\ref{subfig:2PiMass_tbin11}. This background has a $-t$ dependence; it amounts to about 20\% in the low $-t$ region and reduces to about 10\% in the high $-t$ region. 

\subsection{Measurement of SDMEs}
In the beam photon energy range between $8.2$ and $8.8$~GeV used in this analysis, the $t$-channel production process dominates. In order to understand the production mechanism of the $\Delta^{++}(1232)$ recoiling against a $\pi^-$, the spin transfer from the linearly-polarized beam photon to the $\Delta^{++}(1232)$ is studied in terms of the spin-density matrix elements. The SDMEs $\rho^k_{ij}$ describe the spin polarization of the $\Delta^{++}(1232)$, where $i$ and $j$ represent twice the spin projection quantum number of the $\Delta^{++}$, \textit{i.e.} $i/2,j/2 \in (-3/2,-1/2,+1/2,+3/2)$. 

In order to extract the $\Delta^{++}(1232)$ SDMEs, we fit the measured kinematic distribution of the final-state particles with a three-component intensity model:
\begin{equation}
    I_{\text{fit}} =  I_{\text{fit}}^{\text{sig}} +  I_{\text{fit}}^{\pi\pi} +  I_{\text{fit}}^{\text{iso}},
    \label{eq:fitintensity}
\end{equation}
where $I_{\text{fit}}^{\text{sig}}$ describes the intensity distribution of the signal contribution from $\pi^- \Delta^{++}$, $I_{\text{fit}}^{\pi\pi}$ describes the intensity distribution of the di-pion background contribution from $(\pi^- \pi^+) p$, and $I_{\text{fit}}^{\text{iso}}$ describes the background contribution that is isotropically distributed in phase space. These contributions are added incoherently. $I_{\text{fit}}^{\text{sig}}$ and $I_{\text{fit}}^{\pi\pi}$ are described in detail in the sections below.

\subsubsection{$\pi^-\Delta^{++}$ signal parametrization}
The intensity distribution of the signal contribution from $\pi^- \Delta^{++}$ depends on the decay angles of the $\Delta^{++}\rightarrow\pi^+p$. They are studied in the Gottfried-Jackson (GJ) frame, which is defined in the rest frame of the $\Delta^{++}$ with the following coordinate system:   
\begin{equation}
    \hat{z}= \frac{\Vec{p}_{\text{target}}}{|\Vec{p}_{\text{target}}|}, \qquad \hat{y}= \frac{\Vec{p}_\gamma \times \Vec{p}_{\pi^-}}{|\Vec{p}_\gamma \times \Vec{p}_{\pi^-}|} \quad \text{and}  \quad \hat{x}= \hat{y} \times \hat{z},
    \label{eq:coordinates}
\end{equation}
where $\Vec{p}_{\text{target}}$, $\Vec{p}_{\gamma}$, and $\Vec{p}_{\pi^-}$ are the 3-momenta of the target proton, beam photon, and the $\pi^-$, respectively, all in the rest frame of the $\Delta^{++}$. Alternatively, the decay angles of the $\Delta^{++}$ can also be studied in the helicity frame as defined in Eq.~\ref{eq:coordinates_hel}. We discuss our results for the SDMEs in both frames in Sec.~\ref{sec:SDMEs}.

The formalism describing the spin-density matrix elements for a spin-$3/2$ particle is derived and applied for the $K^+\Lambda(1520)$ final state in an earlier publication~\cite{GlueX:2021pcl} and is applied here for the spin-$3/2$ $\Delta^{++}(1232)$. 

The intensity distribution   
\begin{align}
W\left(\theta, \varphi, \Phi\right) = & \frac{3}{4\pi} \Bigl( W^0\left(\theta, \varphi\right) - P_\gamma \cos(2\Phi) W^1\left(\theta, \varphi\right) \nonumber\\
&- P_\gamma \sin(2\Phi) W^2\left(\theta, \varphi\right) \Bigr)
\label{eq:W}\\
W^0\left(\theta, \varphi\right) =& \rho^0_{33}\sin^2\theta + \rho^0_{11}\left(\frac{1}{3} + \cos^2\theta\right)  \nonumber\\ 
&- \frac{2}{\sqrt{3}} \Bigl[\text{Re}(\rho^0_{31})\cos\varphi\sin(2\theta) \nonumber\\
\allowbreak
&+\text{Re}(\rho^0_{3-1})\cos(2\varphi)\sin^2\theta\Bigr] \label{eq:W0}\\
W^1\left(\theta, \varphi\right) =& \rho^1_{33}\sin^2\theta+\rho^1_{11}\left(\frac{1}{3} + \cos^2\theta\right) \nonumber\\
&- \frac{2}{\sqrt{3}}  \Bigl[\text{Re}(\rho^1_{31})\cos\varphi\sin(2\theta) \nonumber \\
&+\text{Re}(\rho^1_{3-1})\cos(2\varphi)\sin^2\theta\Bigr]  \label{eq:W1}\\
W^2\left(\theta, \varphi\right) = & \frac{2}{\sqrt{3}} \Bigl[\text{Im}(\rho^2_{31})\sin\varphi\sin(2\theta)\nonumber \\
&+ \text{Im}(\rho^2_{3-1})\sin(2\varphi)\sin^2\theta \Bigr] 
 \label{eq:W2}
\end{align}
depends on three angles: the polar and azimuthal angle $\theta$, $\varphi$ of the $\pi^+$ in the GJ frame and the angle $\Phi$ between the linear polarization vector of the beam photon and the production plane of the $\Delta^{++}$. The linear polarization of the beam photon provides access to ten SDMEs, four unpolarized ones in Eq.~\ref{eq:W0} and six polarized ones in Eqs.~\ref{eq:W1} and \ref{eq:W2}. To normalize the intensity distribution, the relation $\rho^0_{33} + \rho^0_{11} = \frac{1}{2}$ is used, reducing the number of independent accessible SDMEs to nine. It should be noted that the definition of the $\hat{z}$ axis in Eq.~\ref{eq:coordinates} is defined in the opposite direction to the definition in Ref.~\cite{GlueX:2021pcl}. This leads to a sign change for the SDMEs in Eq.~\ref{eq:W2}.  

In addition to the angular dependence of the intensity, a parameterization for the $\pi^+p$ mass distribution is included in the fit.  The $m_{\pi^+p}$ mass shows clearly the $\Delta^{++}(1232)$ peak. Its mass dependence is described with a relativistic $P$-wave Breit-Wigner function \cite{Breit:1936zzb}
\begin{align}
 \label{eq:BW}
    BW(m_{\pi^+p}) &= \frac{\sqrt{m_0\Gamma_0}}{m_{\pi^+p}^2 - m^2_0 - i m_0\Gamma(m_{\pi^+p}, L)} \quad \text{with}\\
    \Gamma(m_{\pi^+p},L) &= \Gamma_0\frac{q}{m_{\pi^+p}}\frac{m_0}{q_0}\left[\frac{F(q,L)}{F(q_0,L)} \right]^2,
\end{align}
where $F$ is the orbital angular momentum barrier factor as given in Ref.~\cite{VonHippel:1972fg} and $q$ gives the breakup momentum and $q_0$ the breakup momentum at the nominal mass $m_0$. The mass $m_0$ and the width $\Gamma_0$ of the $\Delta^{++}$ are considered as floating fit parameters. The signal intensity model, $I_{\text{fit}}^{\text{sig}}(m_{\pi^+p},\theta,\varphi,\Phi)$, is then given by the product of the angular dependence in Eq.~\ref{eq:W} and the square of the Breit-Wigner amplitude for the mass dependence in Eq.~\ref{eq:BW}.

\subsubsection{Di-pion background parametrization}
As mentioned in Section \ref{sec:eventsel}, the data sample contains an irreducible background stemming from the $\pi^-\pi^+$ meson system. Since this background may contain meson resonances like the $\rho(770)$ and its excited states, its angular distribution cannot be described by phase space, thus a parameterization for the angular dependence is required. We use angular moments to parameterize the angular distribution of the photoproduction of two pseudoscalar mesons ($\pi^-\pi^+$) as given in Ref.~\cite{PhysRevD.100.054017}:
\begin{align}
I(\theta', \varphi', \Phi') =  I^0(\theta', \varphi') &-P_\gamma I^1(\theta', \varphi') \cos(2\Phi')\nonumber \\ 
&-P_\gamma I^2(\theta', \varphi') \sin(2\Phi'). \label{eq:Imoments}
\end{align}
where $\Phi'$ is the angle between the linear polarization vector of the beam photon and the production plane of the proton.  Analogous to the $\pi^+ p$ intensity distribution in Eq.~\ref{eq:W}, the $\pi^-\pi^+$ intensity is decomposed into three different components $I^0, I^1$, and $I^2$, which are given by:
\begin{align}
I^0(\theta', \varphi') = \sum_{L,M\geq0} \left(\frac{2L+1}{4\pi}\right)( &2-\delta_{M0})H^0(LM)\cdot \nonumber\\
&d^L_{M0}(\theta') \cos (M\varphi') \label{eq:I0moments}\\
I^1(\theta', \varphi') = -\sum_{L,M\geq0} \left(\frac{2L+1}{4\pi}\right)( &2-\delta_{M0})H^1(LM)\nonumber \\
&d^L_{M0}(\theta') \cos (M\varphi') \label{eq:I1moments}\\
I^2(\theta', \varphi') = 2\sum_{L,M>0} \left(\frac{2L+1}{4\pi}\right)&\text{Im}H^2(LM)\nonumber \\
&d^L_{M0}(\theta') \sin (M\varphi'),
\label{eq:I2moments}
\end{align}
where $H^\alpha(LM)$ are the moments with $\alpha={0,1,2}$, $d^L_{M0}(\theta')$ is the Wigner $d$-function \cite{wignerd}, and $\theta'$ represents the polar angle and $\varphi'$ the azimuthal angle of $\pi^+$ in the helicity frame of the $\pi^-\pi^+$ system as defined in Ref.~\cite{GlueX:2023fcq}. Truncating the sums in Eqs.~\ref{eq:I0moments}-\ref{eq:I2moments} at different $L_{\text{max}}$ allows us to assess the partial-wave content of the $\pi^-\pi^+$ background contribution. We perform fits with $L_{\text{max}} = {0, 2, 4, 6}$ and find that a good description of the angular distributions, shown in Fig.~\ref{fig_eventsel}, is reached when at least $S$ ($\ell=0$) and $P$ ($\ell=1$) wave contributions are considered. Here, $\ell= 0, 1, 2, 3$ is the total spin of the $\pi^-\pi^+$ system. Including additional moments corresponding to $D$-wave ($\ell=2$) contributions slightly improves the description of the data for certain $-t$ bins, and are therefore also considered in the fit using $L_{\text{max}} = 4$. The effects of truncating at $L_{\text{max}} = 4$ and not including $F$-wave ($\ell=3$) moments are considered in the systematic uncertainty. 

The $\pi^-\pi^+$ mass dependence of the background contribution is parameterized with a Bernstein polynomial of 4th degree in the mass range $[a,b]$, where $a=1.1$~GeV and $b=2.45$~GeV for the nominal fit:
\begin{align}
    \label{eq:Bernstein}
    &B_{4}(m_{\pi^-\pi^+}) = \sum_{i=0}^4 \beta_i b_{i,4}, \\
    &b_{i,4} = \frac{1}{(b-a)^4} \begin{pmatrix}n \\ i \end{pmatrix}(m_{\pi^-\pi^+}-a)^i (b-m_{\pi^-\pi^+})^{4-i}. 
\end{align}
The di-pion background intensity, $I_{\text{fit}}^{\pi\pi}(m_{\pi^-\pi^+},\theta',\varphi',\Phi')$, is then given by the product of Eq.~\ref{eq:Imoments} and Eq.~\ref{eq:Bernstein}.

\subsubsection{Fit method and evaluation}
The data are fit to the model of Eq.~\ref{eq:fitintensity} using the extended maximum likelihood method. The fit method takes into account the detector acceptance based on simulated phase-space events and is explained in more detail in Ref.~\cite{GlueX:2023fcq}. The phase space Monte Carlo sample is generated from 3-body phase space obtained for the $\pi^-\pi^+p$ final state, with acceptance applied using a GEANT4 simulation of the GlueX detector and the same event reconstruction and selection procedures applied as for the data.

In total, the fit model contains 55 fit parameters and 3 external normalization factors: 9 parameters for the $\Delta^{++}$ SDMEs, 2 parameters for the lineshape of $\Delta^{++}$, 5 parameters for the Bernstein polynomial to describe the $\pi^-\pi^+$ mass dependence, and 39 parameters for the $S$, $P$, and $D$-wave moments to parameterize the angular dependence of the $\pi^-\pi^+$ background. The three external normalization factors take into account the three contributions from $\pi^-\Delta^{++}$, $(\pi^-\pi^+) p$, and phase space.

The fit quality is evaluated by comparing the measured distributions to simulated distributions obtained by weighting phase space with the fit results and the detector acceptance. Figure~\ref{fig_eventsel} shows the fit results for two example $t$-bins displaying the $\pi^+p$ and $\pi^-\pi^+$ mass and angular distributions. The yellow curves in Fig.~\ref{fig_eventsel} show the $\pi^-\pi^+$ background contributions in the mass and angular distributions. The background increases towards higher $\pi^+p$ mass and shows a strong $t$-dependence.

\subsection{Statistical and systematic uncertainty}
We determine the statistical uncertainties using the Bootstrapping technique \cite{bootstrap} and proceed in the same way as described in Ref.~\cite{GlueX:2023fcq}.

We consider several contributions to the overall systematic uncertainty of the extracted SDMEs: the chosen event-selection criteria for the $\chi^2/\text{ndf}$ of the kinematic fit and the selections on the $\pi^+p$ and $\pi^-\pi^+$ masses, the accuracy of the $\pi^-\pi^+$ background description, the systematic uncertainty of the degree of linear beam polarization and the sensitivity of the detector system to the four diamond orientations that are used during data-taking.  

The influence of the event selection criteria on the extracted SDMEs is taken into account by varying the selection limits and taking the standard deviation of the different fit results obtained with the varied criteria as an estimation for the systematic uncertainty. For the applied variation of the selections on the $\chi^2/\text{ndf}$ and $m_{\pi^+p}$, the total event sample size is not changed by more than 10\%. However, since the irreducible $\pi^-\pi^+$ background is a significant contribution to the intensity, the lower limit on the $m_{\pi^-\pi^+}$ is increased from $1.1$~GeV up to $1.7$~GeV, which decreases the event sample by almost 40\%. Despite the large change in the sample size, the extracted SDMEs remain stable, indicating that the fit model assumptions and results are robust. The absolute systematic uncertainty on the SDMEs found in this study is of the order of $3 \times 10^{-2}$, which represents the second largest contribution to the systematic uncertainty.

The $\pi^-\pi^+$ background is also studied by varying the degree of the Bernstein polynomial stepwise from 3 to 6. Furthermore, the difference between using $S$, $P$, and $D$-wave moments and using $S, P$, $D$, and $F$-wave moments for the description of the angular distributions of the $\pi^-\pi^+$ system is taken into account for the systematic uncertainty. The SDMEs $\rho^0_{11}$ and $\rho^1_{11}$ are most sensitive to the angular parameterization used for the $\pi^-\pi^+$ background. Both variations account only for a very small part of the overall systematic uncertainty. Instead of using the Bernstein polynomial and angular moments to parameterize the $\pi^-\pi^+$ background, we also perform a study where we describe the mass dependence of the $\pi^-\pi^+$ background using three $\rho$ states, $\rho(770)$, $\rho(1450)$, and $\rho(1700)$ and a phase space contribution, combined with the angular dependence from the $\rho(770)$ SDMEs from Ref.~\cite{GlueX:2023fcq}. This alternative parametrization gives consistent results for the $\Delta^{++}$ SDMEs within the estimated systematic uncertainty.  

A larger contribution to the systematic uncertainty comes from independently fitting the data that correspond to two pairs of orthogonal diamond orientations with their linear polarization direction rotated with respect to the detector system by ($0^\circ, 90^\circ$) and ($45^\circ, -45^\circ$), respectively. The deviation between fits to these two independent datasets is assigned as a systematic uncertainty.  Finally, the linear polarization degree is varied by its systematic uncertainty of $1.5$\% and the difference taken into account for the systematic uncertainty. 

All sources of systematic uncertainty are added in quadrature to obtain the total systematic uncertainty.  

\section{Results}
\subsection{SDMEs}
\label{sec:SDMEs}
\begin{figure*}[htb]
	\centering 
	\includegraphics[width=\textwidth]{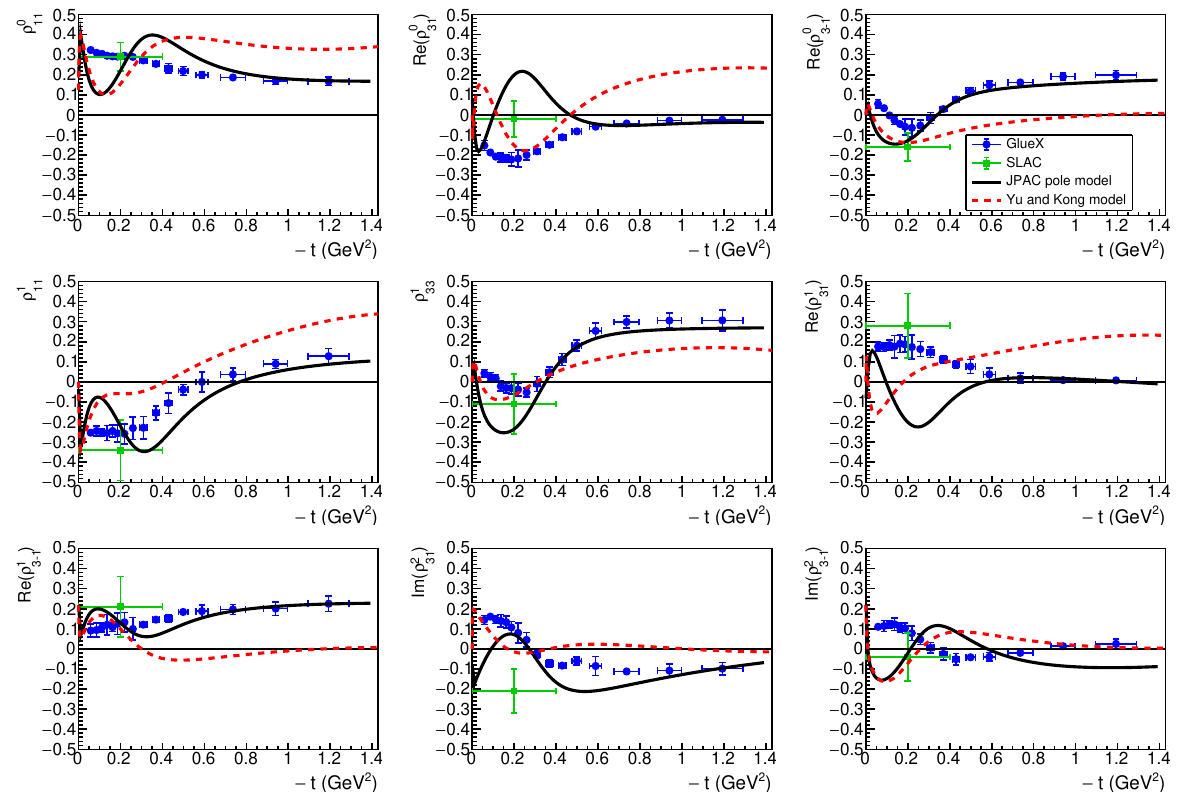}	
	\caption{Spin-density matrix elements of $\Delta^{++}(1232)$ in the Gottfried-Jackson frame as a function of the momentum transfer squared $-t$ from the beam photon to the $\pi^-$. The vertical error bars consist of the statistical and systematic uncertainties added in quadrature. Our data (blue circles) are compared to the previous Ballam et al. measurement (green squares) \cite{PhysRevD.7.3150} and to predictions of the JPAC pole model (solid black line) \cite{nysFeaturesPiDelta2018} and the Yu and Kong model (dashed red line) \cite{YU2017262}. } 
	\label{fig_sdmes}%
\end{figure*}

The SDMEs are measured in the GJ frame for 16 independent $t$ bins and are shown as a function of $-t$ in Fig.~\ref{fig_sdmes}. The vertical error bars consist of both statistical and systematic uncertainties, which are added in quadrature. The combined uncertainty is dominated by the systematic uncertainty for the entire $t$ range. The horizontal position of the data points is given by the mean of the $t$ distribution in the respective $t$ bin and the horizontal error bars are given by the root-mean-square value of the $t$ distribution within that bin. We compare our results to the only previous measurement by Ballam et al.~\cite{PhysRevD.7.3150}, which used $E_\gamma=9.3$~GeV and only one wide $-t\leq 0.4$~GeV$^2$ bin. The Ballam et al. data agree well with our results except for $\text{Re}(\rho^0_{31})$ and $\text{Im}(\rho^2_{31})$, where deviations outside the given uncertainty intervals of the Ballam et al. data are observed. We report our results in much finer bins of $t$, allowing us for the first time to precisely study the $t$ dependence of the SDMEs for $-t < 1.4$~GeV$^2$.

We also compare our results to predictions from the pole model by the JPAC group~\cite{nysFeaturesPiDelta2018}, which is based on Regge-theory amplitudes. The model takes into account natural and unnatural-parity $t$-channel exchange processes, where the latter are described in terms of pseudoscalar meson ($\pi$) and axial-vector meson ($b_1$) exchanges, while the former are described in terms of vector meson ($\rho$) and tensor meson ($a_2$) exchanges. The $t$ dependence of $\rho^1_{11}, ~\rho^1_{33},~\text{Re}(\rho^0_{3-1})$ and $~\text{Re}(\rho^1_{3-1})$ is qualitatively in agreement with the JPAC model predictions, apart from some deviations at small $-t$. However, large discrepancies are visible for the remaining SDMEs. In particular, the model exhibits zero crossings in $\text{Re}(\rho^0_{31})$, $\text{Re}(\rho^1_{31})$, and $\text{Im}(\rho^2_{31})$, which are not observed in the data.  

Since the JPAC model predictions are derived first in the helicity frame and later rotated to the GJ frame, we extract the SDMEs in the helicity frame as well (see Fig.~\ref{fig_sdmes_hel} in the Appendix) and compare them again to the JPAC model. In the helicity frame, the model predictions show a reasonable, qualitative agreement for the shape and magnitude of all SDMEs. However, the $~\text{Re}(\rho^0_{31})$, $~\text{Re}(\rho^1_{31})$ and $~\text{Im}(\rho^2_{31})$ in the helicity frame (Fig.~\ref{fig_sdmes_hel}) have the opposite sign compared to the JPAC pole model predictions. This difference is caused by the relative sign of two helicity amplitude couplings in the JPAC model, which could not be determined from previous measurements of the differential cross section~\cite{PhysRevLett.22.148,PhysRevD.20.1553} and the linearly-polarized beam asymmetry~\cite{PhysRevD.20.1553,PhysRevC.103.L022201}, but they can now be fixed with the current measurements~\cite{JPAC:privcom}. When rotating the SDMEs to the GJ frame, this sign ambiguity affects all the SDMEs and makes the comparison of the JPAC model to the data more challenging.

The model by Yu and Kong \cite{YU2017262} is based on Regge theory as well and considers the $t$-channel $\pi$, $\rho$, and $a_2$ meson exchanges. According to their model, the $\pi^-\Delta^{++}$ production mechanism is dominated by $\pi$ exchange and the tensor meson $a_2$ is found to play a more important role than the $\rho$ meson exchange in order to describe the previous differential cross section and beam asymmetry data. However, the $b_1$ meson exchange is found to not be needed and therefore not included in their model. Overall, the $t$ dependence of the SDMEs is not well described by the Yu and Kong model.

To further investigate the deviations between our data and the model predictions and to separate the unnatural-parity and natural-parity exchanges, linear combinations of SDMEs are discussed in the following section.

\subsection{SDMEs for natural and unnatural-parity exchange}
\begin{figure*}[htb]
	\centering 
	\includegraphics[width=\textwidth]{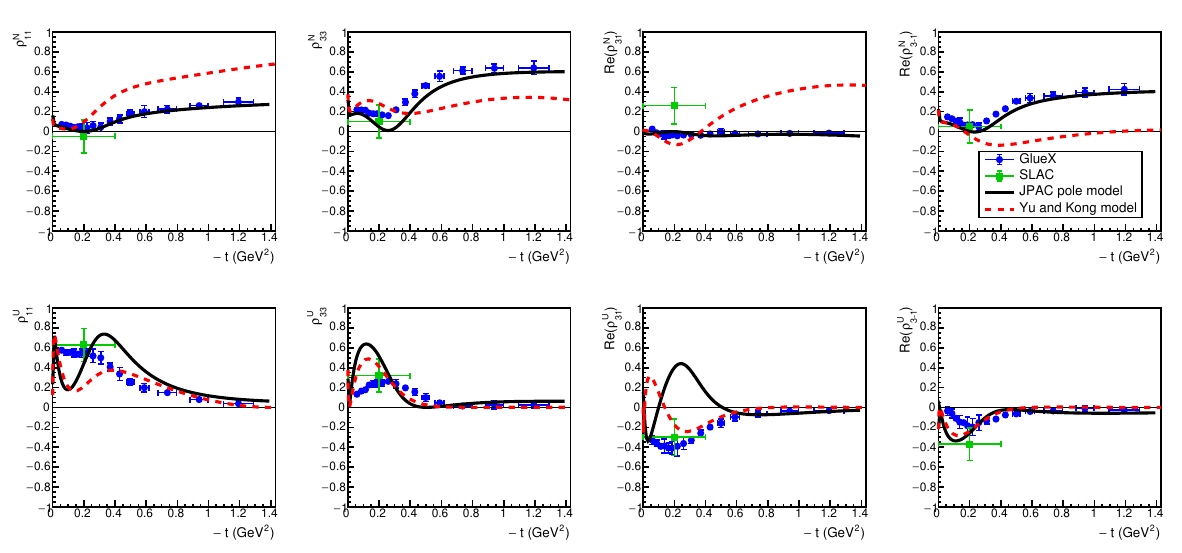}	
    \caption{Linear combinations of SDMEs (see Eq.~\ref{eq:sdmecombos}) that represent natural (N, top row) and unnatural (U, bottom row) exchange components. Our data (blue circles) are compared to the previous Ballam et al. measurement (green squares) \cite{PhysRevD.7.3150} and to predictions of the JPAC pole model (solid black line) \cite{nysFeaturesPiDelta2018} and the Yu and Kong model (dashed red line) \cite{YU2017262}. }
	\label{fig_sdmecombos}%
\end{figure*}

We construct linear combinations of the extracted SDMEs to decompose the SDMEs into the unnatural-parity (U) and natural-parity (N) exchange components:
\begin{equation}
    \rho^{\text{N}/\text{U}}_{ij} = \rho^{0}_{ij} \pm \rho^{1}_{ij}.
    \label{eq:sdmecombos}
\end{equation}
The relations between $\rho^{\text{N}/\text{U}}_{ij}$ and the unnatural and natural-exchange amplitudes are given in Ref.~\cite{GlueX:2021pcl}. Figure~\ref{fig_sdmecombos} shows in the top row the natural and in the bottom row the unnatural-exchange SDME components. They are compared again to the JPAC pole model and the Yu and Kong model predictions. Comparing the top and bottom rows, it becomes evident that in the low $-t$ region up to about $0.45$~GeV$^2$ unnatural-parity exchange dominates since in this region the natural contributions $\rho^{\text{N}}_{ij}$ are close to zero, while the unnatural contributions $\rho^{\text{U}}_{ij}$ are large. The situation is reversed for the $-t$ region above $0.45$~GeV$^2$, where natural-parity exchange dominates. The comparison to the JPAC model reveals that the natural-parity exchange is well modeled by the JPAC pole model. The large deviations between the data and the JPAC pole model, that are visible in the SDMEs in Fig.~\ref{fig_sdmes}, can be traced back to the unnatural-parity exchange component, where $\pi$ exchange is assumed to dominate in the $t$-channel process. 

The Yu and Kong model describes the natural-parity exchange well for very low $-t$ values below $0.3$~GeV$^2$, but cannot predict the $t$ dependence of the data for larger $-t$ values and either overestimates (for $\rho^\text{N}_{11}$ and $\text{Re}(\rho^\text{N}_{31})$) or underestimates (for $\rho^\text{N}_{33}$ and $\text{Re}(\rho^\text{N}_{3-1})$) the natural-parity contributions. The opposite is the case for unnatural parity-exchange, where the Yu and Kong model does not predict the data well for $-t$ values below $0.3$~GeV$^2$, but reproduces the data well for higher $-t$ values.

\subsection{Beam asymmetry}
\begin{figure}[h!]
	\centering 
	\includegraphics[width=\columnwidth]{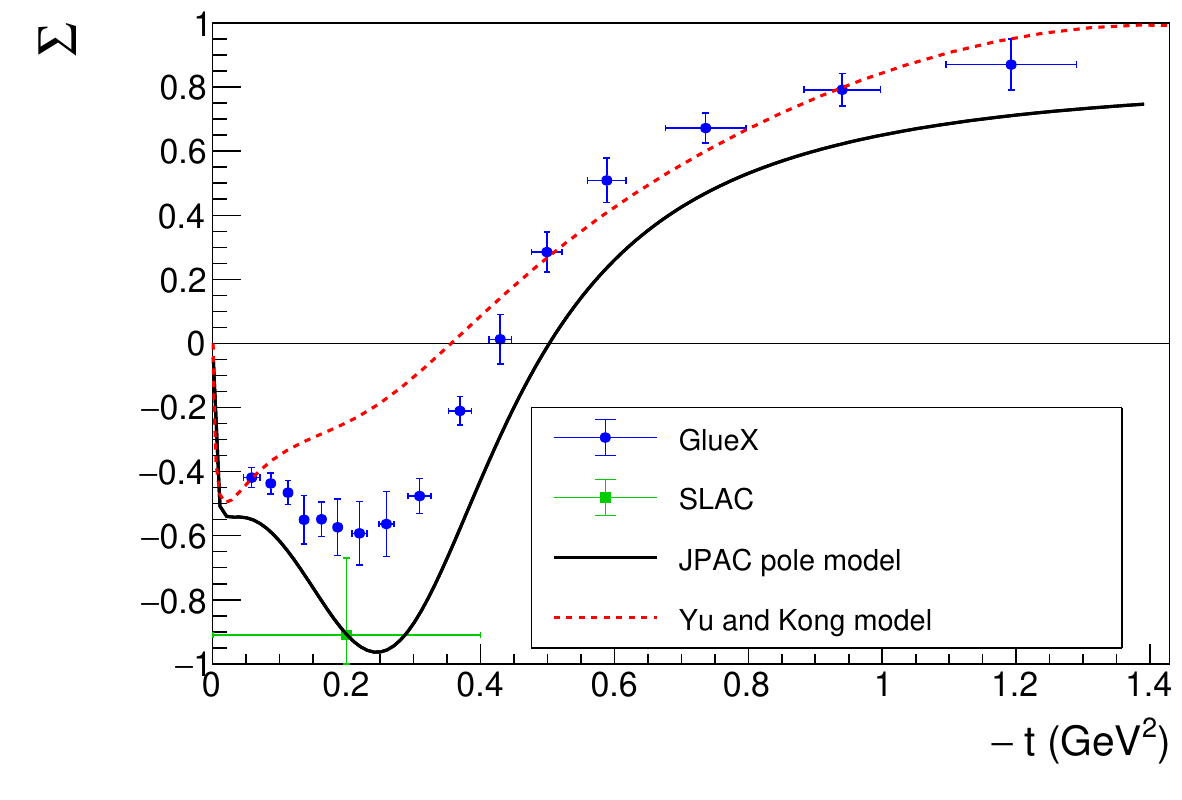}	
    \caption{Beam asymmetry extracted from the SDMEs using Eq.~(\ref{eq:sigma}) (blue circles). The GlueX data are compared to the previous Ballam et al. data (green squares) \cite{PhysRevD.7.3150} and to predictions of the JPAC pole model (solid black line) \cite{nysFeaturesPiDelta2018} and the Yu and Kong model (dashed red line) \cite{YU2017262}. }
	\label{fig_sigma}%
\end{figure}

Using the sum of the two SDMEs $\rho^1_{11}$ and $\rho^1_{33}$, the beam asymmetry $\Sigma$ is determined as follows:
\begin{equation}
    \Sigma = 2\left(\rho^1_{11} + \rho^1_{33}\right).
    \label{eq:sigma}
\end{equation}
The results for the beam asymmetry are shown in Fig.~\ref{fig_sigma} and supersede our previously published results~\cite{PhysRevC.103.L022201}. The previous results used only 17\% of the data analyzed in this work and were extracted by measuring the asymmetry of the event yield for two perpendicular diamond settings and integrating over the $\Delta^{++}$ decay angles $\theta$ and $\varphi$. 
A comparison of the two results is given in Fig.~\ref{fig_sigma_comp} and the two different extraction methods are discussed in detail in ~\ref{sec:beamasym_appendix}.

The JPAC pole model describes the shape of the $t$ dependence of $\Sigma$ qualitatively well, but shows large deviations in the magnitude of the beam asymmetry especially in the low $-t$ range. The Yu and Kong model describes the beam asymmetry well in the range above $-t=0.5$~GeV$^2$, but does not predict the dip at $-t=0.25$. 

\section{Summary and conclusions}
\label{sec:summary}
We present measurements of the $\Delta^{++}(1232)$ SDMEs for the photoproduction reaction $\gamma p \to \pi^-\Delta^{++}$ that are obtained using the GlueX detector with a linearly-polarized photon beam from $8.2$ to $8.8$~GeV. Our measurements constitute the first precise determination of the $t$ dependence of the $\Delta^{++}(1232)$ SDMEs. A comparison to the JPAC and Yu and Kong Regge-theory models shows that the $t$-dependence of the SDMEs is not well reproduced, although previous unpolarized cross section and beam asymmetry data of Ballam et al.~\cite{PhysRevD.7.3150} are described by the models adequately. The SDMEs are sensitive to the relative sign of the model couplings, which was undetermined from previous data. Separating the natural-parity and unnatural-parity exchange components through linear combinations of SDMEs reveals that the deviations between data and the JPAC Regge-theory model can be traced back to the unnatural-parity exchange component.
Thus, our data provide important constraints on the Regge-theory models, specifically on the unnatural-parity pion exchange, which is expected to play an important role in many other charge-exchange photoproduction reactions to be studied in the search for exotic mesons at GlueX. In particular, the charge-exchange reaction $\gamma p \to \eta'\pi^-\Delta^{++}$ is crucial in attempting to confirm the existence of the lightest hybrid meson $\pi_1(1600)$ based on an estimation of the upper limit for the $\pi_1(1600)$ photoproduction cross section \cite{Afzal:2024ulu}. 

\section*{Acknowledgements}
We thank Adam Szczepaniak, Vincent Mathieu, Vanamali Shastry, and Gloria Montaña from the JPAC group and B.-G. Yu for the fruitful discussions. 
The analysis in this article was supported by the U.S. Department of Energy, Office of Science, Office of Nuclear Physics under contract DOE Grant No. DE-FG02-87ER40315. The work of F. Afzal is supported by the Argelander Mobility Grant awarded by the University of Bonn and J. R. Stevens is supported by DOE Grant DE-SC0023978. We would like to acknowledge the outstanding efforts of the staff of the Accelerator and the Physics Divisions at Jefferson Lab that made the experiment possible. This work was also supported in part by the U.S. Department of Energy, the U.S. National Science Foundation, NSERC Canada, the German Research Foundation, GSI Helmholtzzentrum f\"ur Schwerionenforschung GmbH, the Russian Foundation for Basic Research, the UK Science and Technology Facilities Council, the Chilean Comisión Nacional de Investigación Científica y Tecnológica, the National Natural Science Foundation of China, and the China Scholarship Council. This material is based upon work supported by the U.S. Department of Energy, Office of Science, Office of Nuclear Physics under contract DE-AC05-06OR23177.  
This research used resources of the National Energy Research Scientific Computing Center (NERSC), a U.S. Department of Energy Office of Science User Facility operated under Contract No. DE-AC02-05CH11231. This work used the Extreme Science and Engineering Discovery Environment (XSEDE), which is supported by National Science Foundation grant number ACI-1548562. Specifically, it used the Bridges system, which is supported by NSF award number ACI-1445606, at the Pittsburgh Supercomputing Center (PSC).

\appendix
\section{SDMEs in the helicity frame}
\begin{figure*}[htb]
	\centering 
	\includegraphics[width=\textwidth]{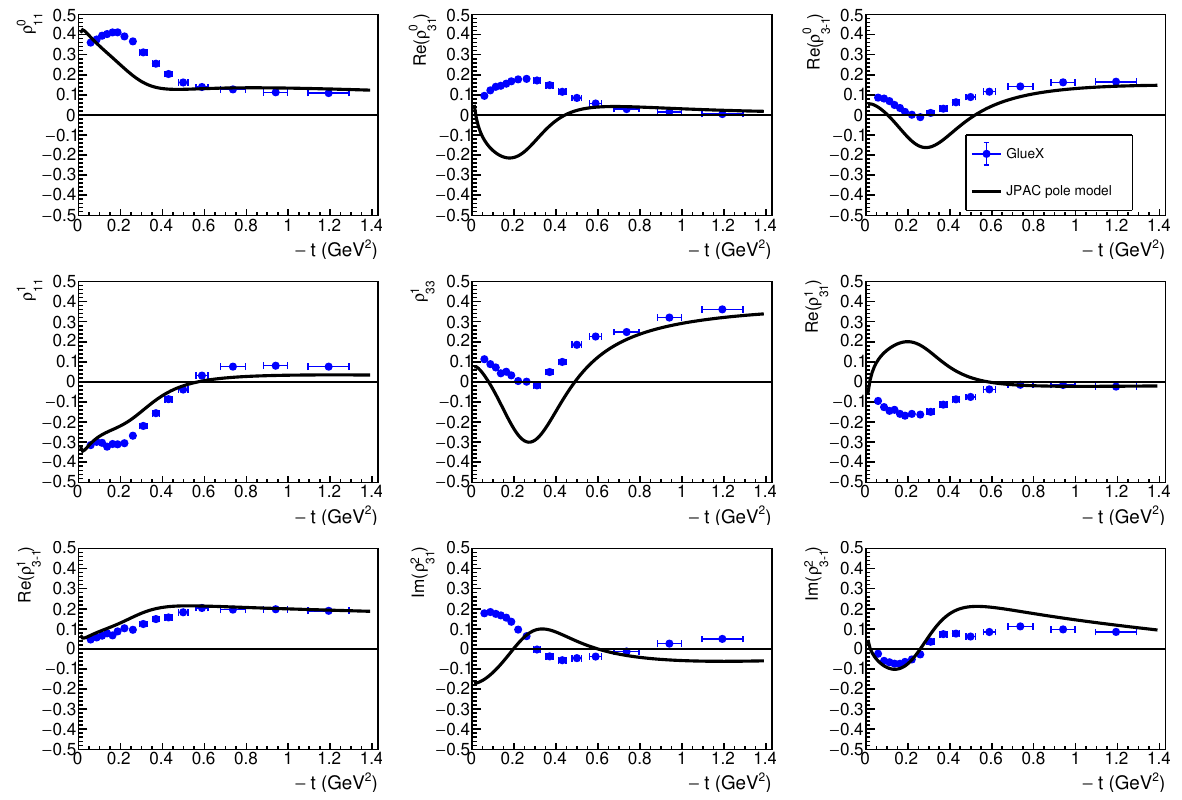}	
	\caption{Spin-density matrix elements of $\Delta^{++}(1232)$ in the helicity frame. They are shown as a function of the momentum transfer squared $-t$ from the incoming photon to the $\pi^-$. The vertical error bars consist of only the statistical uncertainties. Our data (blue circles) are compared to predictions of the JPAC pole model (solid black line) \cite{nysFeaturesPiDelta2018}. } 
	\label{fig_sdmes_hel}%
\end{figure*}

Fig.~\ref{fig_sdmes_hel} shows the $\Delta^{++}(1232)$ SDMEs in the helicity frame, which is defined by the following coordinate system in the rest frame of the $\Delta^{++}$:
\begin{equation}
    \hat{z}= \frac{-\Vec{p}_{\pi^-}}{|\Vec{p}_{\pi^-}|}, \qquad \hat{y}= \frac{\Vec{p}_\gamma \times \Vec{p}_{\pi^-}}{|\Vec{p}_\gamma \times \Vec{p}_{\pi^-}|}\quad \text{and}  \quad \hat{x}= \hat{y} \times \hat{z}.
    \label{eq:coordinates_hel}
\end{equation}

\section{Beam asymmetry - Comparison of GlueX data}
\label{sec:beamasym_appendix}
\begin{figure}[h!]
	\centering 
	\includegraphics[width=\columnwidth]{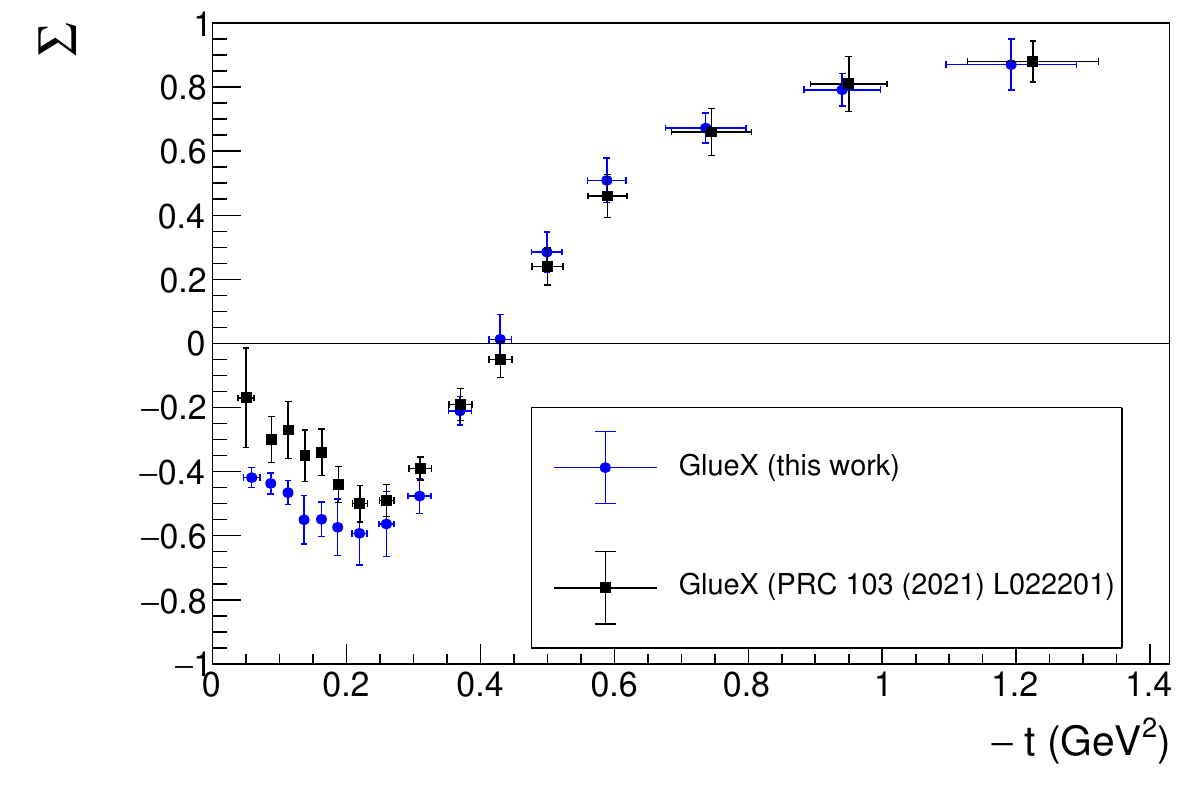}	
    \caption{Beam asymmetry extracted from the SDMEs using Eq.~(\ref{eq:sigma}) (blue circles), and our previously published results based on the yield-asymmetry method (black circles)~\cite{PhysRevC.103.L022201}. Note: The latter data points are shown at the $t$-bin centers, while for the present results the mean value of the $t$ distribution for each $t$-bin is used. }
	\label{fig_sigma_comp}%
\end{figure}

Fig.~\ref{fig_sigma_comp} shows the results for the beam asymmetry extracted from the SDMEs (see Eq.~\ref{eq:sigma}). These results are compared with our previously published results~\cite{PhysRevC.103.L022201}, where the beam asymmetry is determined from the asymmetry of the event yield for two perpendicular diamond settings and by integrating over the $\Delta^{++}$ decay angles. In general, both results show a strong $t$ dependence for the beam asymmetry and are in good agreement over a large $-t$ range. However, in the very low $-t$ range, a discrepancy is visible between the two methods. The yield-asymmetry method has the disadvantage that it can lead to biased results for $\Sigma$ in case of a non-uniform detection efficiency of the $\Delta^{++}$ decay angles. This bias stemming from the efficiency modeling was included in the systematic uncertainty in the previous publication. This uncertainty was estimated using Monte Carlo simulation based on the difference between JPAC models for the SDMEs~\cite{nysFeaturesPiDelta2018}. However, the uncertainty seems to have been underestimated at low $-t$. In this region, the JPAC model fails to reproduce well the SDMEs that enter Eq.~\ref{eq:sigma} while at the same time the detection efficiency is highly non-uniform in $\cos\theta$ (see Fig.~\ref{subfig:cosT_tbin2}). Our present SDME extraction method for the beam asymmetry according to Eq.~\ref{eq:sigma} does not have any such bias since the full $\Delta^{++}$ decay angular phase space is analyzed. Finally, we confirm this bias by analyzing Monte Carlo simulation, generated with the measured SDME values, and by extracting the beam asymmetry from this simulation sample using the yield-asymmetry method. This reproduces the biased beam asymmetry values from our previously published results~\cite{PhysRevC.103.L022201}.

\bibliographystyle{elsarticle-num-names} 
%
\bibliography{main}

\end{document}